\documentclass[12pt,a4paper]{article}
\usepackage{amssymb}
\usepackage{graphicx}
\usepackage{amsmath}
\usepackage{makeidx}
\usepackage{indentfirst}
\usepackage[T1]{fontenc}
\usepackage[utf8]{inputenc}
\usepackage{authblk}

\newcounter{resultnum}[section]
\newcounter{conclusionnum}[section]
\newcounter{conditionnum}[section]
\newcounter{conjecturenum}[section]
\newcounter{examplenum}[section]
\newcounter{exercisenum}[section]
\newcounter{lemmanum}[section]
\newcounter{notationnum}[section]
\newcounter{theoremnum}[section]
\newcounter{definitionnum}[section]
\newcounter{corollarynum}[section]
\newcounter{remarknum}[section]
\newcounter{propositionnum}[section]
\newcounter{acknowledgementnum}[section]
\newcounter{algorithmnum}[section]
\newcounter{axiomnum}[section]
\newcounter{casenum}[section]
\newcounter{claimnum}[section]
\newcounter{summarynum}[section]
\newcounter{problemnum}[section]
\begin{document}

\title{Off--Diagonal Ekpyrotic Scenarios and Equivalence of Modified,
Massive and/or Einstein Gravity}
\date{November 09, 2015}
\author{ \vspace{.1 in} {\large \textbf{Sergiu I. Vacaru}} \\
\vspace{.1 in} {\small \textit{University "Al. I. Cuza" Ia\c si, Rector's
Department }}\\
{\small \textit{14 Alexandru Lapu\c sneanu street, Corpus R, UAIC, office 323, Ia\c si, Romania 700057; }}\\
{\small and\thanks{visiting DAAD fellowship with two affiliations in Germany, Oct 1 -December 31, 2015} }\\
{\small \textit{Max-Planck-Institute for Physics, Werner-Heisenberg-Institute, }}\\
{\small Foehringer Ring 6, M\"{u}nchen, Germany D-80805 ;}\\
{\small and}\\
{\small Riemann Center for Geometry and Physics, \textit{Leibniz Universit\"{a}t Hannover}}\\
{\small Welfengarten 1, 30167 Hannover, Germany}\\
{\small \textit{email: sergiu.vacaru@uaic.ro; sergiu.vacaru@gmail.com }}}
\maketitle

\begin{abstract}
Using our anholonomic frame deformation method, we show how generic off--diagonal cosmological solutions depending, in general, on all spacetime coordinates and undergoing a phase of ultra--slow contraction  can be constructed in massive gravity.  In this paper, there are found and studied new classes
of locally anisotropic and (in) homogeneous cosmological metrics with open
and closed spatial geometries. The late time
acceleration is present  due to effective cosmological terms induced by nonlinear
off--diagonal interactions and graviton mass. The off--diagonal cosmological metrics and
related St\" uckelberg fields are constructed in explicit form up to
nonholonomic frame transforms of the Friedmann--Lama\^{\i}%
tre--Robertson--Walker (FLRW) coordinates. We show that the solutions include matter,
graviton mass and other effective sources modelling  nonlinear gravitational
and matter fields interactions in modified and/or massive gravity, with polarization of physical constants and
deformations of metrics, which may explain certain dark energy and dark
matter effects. There are stated and analyzed the conditions when such configurations
mimic interesting solutions in general relativity and modifications and
recast the general Painlev\' e--Gullstrand and FLRW metrics. Finally, we elaborate on  a reconstruction procedure for a subclass of off--diagonal
cosmological solutions which describe cyclic and ekpyrotic universes, with an emphasis on open issues and observable signatures.

\vskip5pt

 \vskip5pt {\small %
}
\end{abstract}





The idea that graviton may have a nontrivial mass was proposed by Fierz and
Pauli work \cite{fierz} (for recent reviews and related f(R) modifications,
see \cite{odints}). The key steps in elaborating a modern version of a ghost
free (bimetric) massive gravity theory were made in a series of papers: The
so--called vDVZ discontinuity problem was solved using the Vainshtein
mechanism \cite{dvz} (avoiding discontinuity by going beyond the linear
theory), or following more recent approaches based on DGP model \cite{dgp}.
But none solution was found for another problem with ghosts because at
nonlinear order in massive gravity appears a sixth scalar degree of freedom
as a ghost (see the Boulware and Deser paper and similar issues related to
the effective field theory approach in Refs. \cite{boul}). That stoppled for
almost two decades the research on formulating a consistent theory of
massive gravity.

Recently, a substantial progress was made when de Rham and co--authors had
shown how to eliminate the scalar mode and Hassan and Rosen established a
complete proof for a class of bigravity / bimetric gravity theories, see
\cite{drg}. The second metric describes an effective exotic matter related
to massive gravitons and does not suffer from ghost instability to all
orders in a perturbation theory and away from the decoupling limit.

The possibility that the graviton has a nonzero mass $\mathring{\mu}$
results not only in fundamental theoretical implications but give rise to
straightforward phenomenological consequences. For instance, a gravitational
potential of Yuka\-wa form $\sim e^{-\mathring{\mu}r}/r$ results in decay of
gravitational interactions at scales $r\geq \mathring{\mu}^{-1}$ and this
can result in the accelerated expansion of the Universe. This way, a theory
of massive gravity provides alternatives to dark energy and, via effective
polarizations of fundamental physical constants (in result of generic
off--diagonal nonlinear interactions), may explain certain dark matter
effects. Recently, various cosmological models derived for ghost free
(modified) massive gravity and bigravity theories have been elaborated and
studied intensively (see, for instance, Refs. \cite{mgcosm,odints,kobayashi}%
).

The goal of this work is to construct generic off--diagonal cosmological
solutions in massive gravity theory, MGT, and state the conditions when such
configurations are modelled equivalently in general relativity (GR). We
shall develop and apply in massive gravity theory the so--called anholonomic
frame deformation method, AFDM \cite{vmgpapers}. As a first step, we
consider off--diagonal deformations of a "prime" cosmological solution taken
in general Painlev\'{e}--Gullstrand (PG) form, when the Friedman--Lama\^{\i}%
tre--Robertson--Walker (FLRW) can be recast for well--defined geometric
conditions. At the second step, the "target" metrics will be generated to
possess one Killing symmetry (or other none Killing symmetries) and depend
on timelike and certain (all) spacelike coordinates. In general, such
off--diagonal solutions are with local anisotropy and inhomogeneities for
effective cosmological constants and polarizations of other physical
constants and coefficients of cosmological metrics which can be modelled
both in MGT and GR. Finally (the third step), we shall emphasize and
speculate on importance of off--diagonal nonlinear gravitational
interactions for elaborating cosmological scenarios with anisotropic
polarization of vacuum and/or de Sitter like configurations.

We study modified massive gravity theories determined on a
pseudo--Riemannian spacetime $\mathbf{V}$ with physical metric $\mathbf{g=\{g%
}_{\mu \nu }\}$ and certain fiducial metric as we shall explain below. The
action for our model is
\begin{eqnarray}
S &=&\frac{1}{16\pi }\int \delta u^{4}\sqrt{|\mathbf{g}_{\alpha \beta }|}[%
\widehat{f}(\widehat{R})-\frac{\mathring{\mu}^{2}}{4}\mathcal{U}(\mathbf{g}%
_{\mu \nu },\mathbf{K}_{\alpha \beta })+\ ^{m}L]  \label{act} \\
&=&\frac{1}{16\pi }\int \delta u^{4}\sqrt{|\mathbf{g}_{\alpha \beta }|}%
[f(R)+\ ^{m}L].  \label{act1}
\end{eqnarray}%
In this formula, $\widehat{R}$ is the scalar curvature for an auxiliary
(canonical) connection $\widehat{\mathbf{D}}$ uniquely determined by two
conditions 1) it is metric compatible, $\widehat{\mathbf{D}}\mathbf{g}=0%
\mathbf{,}$ and 2) its $h$- and $v$-torsions are zero (but there are nonzero
$h-v$ components of torsion $\widehat{\mathcal{T}}$ completely determined by
$\mathbf{g)}$ for a conventional splitting $\mathbf{N}:\ T\mathbf{V=}h%
\mathbf{V\oplus }v\mathbf{V,}$ see details in \cite{vadm1}.\footnote{%
We consider a conventional $2+2$ splitting when coordinates are labelled in
the form $u^{\alpha }=(x^{i},y^{a}),$ or $u=(x,y),$ with indices $%
i,j,k...=1,2$ and $a,b,...=3,4.$ There will be used boldface symbols in
order to emphasize that certain geometric/physical objects and/or formulas
are written with respect to N--adapted bases (\ref{nframes}). There will be
considered also left up/low indices as labels for some geometric/physical
objects.} The "priority" of the connection $\widehat{\mathbf{D}}$ is that it
allows to decouple the field equations in various gravity theories and
construct exact solutions in very general forms. We shall work with generic
off--diagonal metrics and generalized connections depending on all spacetime
coordinates when, for instance, of type $\widehat{\mathbf{D}}=\nabla +%
\widehat{\mathbf{Z}}[\widehat{\mathcal{T}}].$ Such distortion relations from
the Levi--Civita (LC) connection $\nabla $ are uniquely determined by a
distorting tensor $\widehat{\mathbf{Z}}$ completely defined by $\widehat{%
\mathcal{T}}$ and (as a consequence for such models) by $(\mathbf{g},\mathbf{%
N}).$ Having constructed certain general classes of solutions (for instance,
locally anisotropic and/or inhomogeneous cosmological ones), we can impose
additional nonholonomic (non--integrable constraints) when $\widehat{\mathbf{%
D}}_{\mid \widehat{\mathcal{T}}=0}\rightarrow \nabla $ and $\widehat{R}%
\rightarrow R,$ where $R$ is the scalar curvature of $\nabla ,$ and it is
possible to extract exact solutions in GR.

The theories with actions of type (\ref{act}) generalize the so--called
modified $f(R)$ gravity, see reviews and original results in \cite{odints},
and the ghost--free massive gravity (by de Rham, Gabadadze and Tolley, dRGT)
\cite{drg}. We use the units when $\hbar =c=1$ and the Planck mass $M_{Pl}$
is defined via $M_{Pl}^{2}=1/8\pi G$ with 4--d Newton constant $G.$ We write
$\delta u^{4}$ instead of $d^{4}u$ because there are used N--elongated
differentials (see below the formulas (\ref{nframes})) and consider $%
\mathring{\mu}=const$ as the mass of graviton. For LC--configurations, we
can fix (as particular cases) conditions of type
\begin{equation}
\widehat{f}(\widehat{R})-\frac{\mathring{\mu}^{2}}{4}\mathcal{U}(\mathbf{g}%
_{\mu \nu },\mathbf{K}_{\alpha \beta })=f(\widehat{R}),\mbox{ \ or \ }%
\widehat{f}(\widehat{R})=f(R),\mbox{ \ or \ }\widehat{f}(\widehat{R})=R,
\label{mgrfunct}
\end{equation}%
which depend on the type of models we elaborate and what classes of
solutions we wont to construct. The first one is necessary if we wont to
encode massive gravity effects into a MGT with a generalized connection and
corresponding Ricci scalar $\widehat{R}$ which allows us to decouple the
gravitational field equations and generate off--diagonal solutions. The
second condition is necessary for extracting MGT models with Levi--Civita
conditions. We can also consider the third type classes of solutions when
theories with both $f$-- and linear connection modifications are effectively
modelled as certain off--diagonal solutions in GR. It will be possible to
find solutions in explicit form if we fix the coefficients $\{N_{i}^{a}\}$
of $\mathbf{N}$ and local frames for $\widehat{\mathbf{D}}$ when $\widehat{R}%
=const$ and $\partial _{\alpha }\widehat{f}(\widehat{R})=(\partial _{%
\widehat{R}}\widehat{f})\times \partial _{\alpha }\widehat{R}=0$ but, in
general, $\partial _{\alpha }f(R)\neq 0.$ The equations of motion for such
modified massive gravity theory can be written\footnote{%
see details on action and variational methods in \cite{drg}; we shall follow
some conventions from \cite{kobayashi}; the Einstein summation rule on
repeating indices will be applied if the contrary is not stated}%
\begin{equation}
(\partial _{\widehat{R}}\widehat{f})\widehat{\mathbf{R}}_{\mu \nu }-\frac{1}{%
2}\widehat{f}(\widehat{R})\mathbf{g}_{\mu \nu }+\mathring{\mu}^{2}\mathbf{X}%
_{\mu \nu }=M_{Pl}^{-2}\mathbf{T}_{\mu \nu },  \label{mgrfe}
\end{equation}%
where $M_{Pl}$ is the Plank mass, $\widehat{\mathbf{R}}_{\mu \nu }$ is the
Einstein tensor for a pseudo--Riemannian metric $\mathbf{g}_{\mu \nu }$ and $%
\widehat{\mathbf{D}},$ $\mathbf{T}_{\mu \nu }$ is the standard matter
energy--momentum tensor. \ For $\widehat{\mathbf{D}}\rightarrow \nabla ,$ we
get $\widehat{\mathbf{R}}_{\mu \nu }\rightarrow R_{\mu \nu }$ with a
standard Ricci tensor $R_{\mu \nu }$ for $\nabla .$ The effective
energy--momentum tensor $\mathbf{X}_{\mu \nu }$ is defined in a
"sophisticate" form by the potential of graviton $\mathcal{U}=\mathcal{U}%
_{2}+\alpha _{3}\mathcal{U}_{3}+\alpha _{4}\mathcal{U}_{4}$, where $\alpha
_{3}$ and $\alpha _{4}$ are free parameters. The values $\mathcal{U}_{2},%
\mathcal{U}_{3}$ and $\mathcal{U}_{4}$ are certain polynomials on traces of
some other polynomials of a matrix $\mathcal{K}_{\mu }^{\nu }=\delta _{\mu
}^{\nu }-\left( \sqrt{g^{-1}\Sigma }\right) _{\mu }^{\nu }$ for a tensor
determined by four St\"{u}ckelberg fields $\phi ^{\underline{\mu }}$ as
\begin{equation}
\Sigma _{\mu \nu }=\partial _{\mu }\phi ^{\underline{\mu }}\partial _{\nu
}\phi ^{\underline{\nu }}\eta _{\underline{\mu }\underline{\nu }},
\label{bm}
\end{equation}%
when $\eta _{\underline{\mu }\underline{\nu }}=(1,1,1,-1).$ Following a
series of arguments presented in \cite{kobayashi}, when the parameter choice
$\alpha _{3}=(\alpha -1)/3,\alpha _{4}=(\alpha ^{2}-\alpha +1)/12$ is useful
for avoiding potential ghost instabilities, we can fix
\begin{equation}
\mathbf{X}_{\mu \nu }=\alpha ^{-1}\mathbf{g}_{\mu \nu }.  \label{cosmconst}
\end{equation}%
De Sitter solutions for an effective cosmological constant are possible, for
instance, for ansatz of PG type,%
\begin{equation}
ds^{2}=U^{2}(r,t)[dr+\epsilon \sqrt{f(r,t)}dt]^{2}+\widetilde{\alpha }%
^{2}r^{2}(d\theta ^{2}+\sin ^{2}\theta d\varphi ^{2})-V^{2}(r,t)dt^{2}.
\label{pgm}
\end{equation}%
In above formula, there are used spherical coordinates labelled in the form $%
u^{\beta }=(x^{1}=r,x^{2}=\theta ,y^{3}=\varphi ,y^{4}=t),$ the function $f$%
\ takes non--negative values and the constant $\widetilde{\alpha }=\alpha
/(\alpha +1)$ and $\epsilon =\pm 1.$ For such bimetric configurations, the St%
\"{u}ckelberg fields are parameterized in the unitary gauge as $\phi ^{%
\underline{4}}=t$ and $\phi ^{\underline{1}}=r\widehat{n}^{\underline{1}%
},\phi ^{\underline{2}}=r\widehat{n}^{\underline{2}},\phi ^{\underline{3}}=r%
\widehat{n}^{\underline{3}},$ where a three dimensional (3--d) unit vector
is defined as $\widehat{n}=(\widehat{n}^{\underline{1}}=\sin \theta \cos
\varphi ,\widehat{n}^{\underline{2}}=\sin \theta \sin \varphi ,\widehat{n}^{%
\underline{3}}=\cos \theta ).$ Any PG metric of type (\ref{pgm}) defines
solutions both in GR and in MGT. It allows us to extract the de Sitter
solution, in the absence of matter, and to obtain standard cosmological
equations with FLRW metric, for a perfect fluid source%
\begin{equation}
T_{\mu \nu }=\left[ \rho (t)+p(t)\right] u_{\mu }u_{\nu }+p(t)g_{\mu \nu },
\label{memt}
\end{equation}%
where $u_{\mu }=(0,0,0,-V)$ can be reproduced for the effective cosmological
constant $\ ^{eff}\lambda =\mathring{\mu}^{2}/\alpha .$ It is also possible
to express metrics of type (\ref{pgm}) in a familiar cosmological FLRW form
(see formulas (23), (24) and (27) in \cite{kobayashi}).

Let us consider an ansatz
\begin{eqnarray}
ds^{2} &=&\eta _{1}(r,\theta )\mathring{g}_{1}(r)dr^{2}+\eta _{2}(r,\theta )%
\mathring{g}_{2}(r)d\theta ^{2}  \label{offdans} \\
&&+\omega ^{2}(r,\theta ,\varphi ,t)\{\eta _{3}(r,\theta ,t)\mathring{h}%
_{3}(r,\theta )[d\varphi +n_{i}(r,\theta )dx^{i}]^{2}  \notag \\
&& +\eta _{4}(r,\theta ,t)\mathring{h}_{4}(r,\theta ,t)[dt+(w_{i}(x^{k},t)+%
\mathring{w}_{i}(x^{k}))dx^{i}]^{2}\},  \notag
\end{eqnarray}%
with Killing symmetry on $\partial _{3}=\partial _{\varphi },$ which (in
general) can not be diagonalized by coordinate transforms. The values $\eta
_{\alpha }$ are called "polarization" functions; $\omega $ is the so--called
"vertical", v, conformal factor.\footnote{%
Up to certain frame/coordinates transforms, any inhomogeneous and/or locally
anistropic cosmological metrics with such a Killing type space symmetry on $%
\partial _{\varphi }$ can be parameterized in a form (\ref{offdans}). This
allows us to deform nonholonomically any class of "prime" cosmological
metrics $\mathring{\mathbf{g}}$, with $\eta _{\beta}=1$ and N--connection
coefficients $\mathring{n}_{i}(x^{k})=0$ and nontrivial $\mathring{w}%
_{i}(x^{k})$, into certain new classes of exact cosmological solutions
defining "target" metrics ${\mathbf{g}}$ with nontrivial $\eta
_{\beta}(x^{k},t)$ and ${n}_{i}(x^{k})$ and ${w}_{i}(x^{k},t)$. For
instance, we can chose a "prime" $\mathring{\mathbf{g}}$ to define a FLRW
diagonal cosmological model and analyze how possible MGT, massive gravity or other type effects  would deform it into a locally anisotropic
inhomogeneous "target" one ${\mathbf{g}}(x^k,t)$. It is possible to generate
target cosmological metrics depending only on a time like coordinate and with very
small off--diagonal deformations which may explain recent observational
cosmological data. Nevertheless, certain nonlinear parametric effects can be
important ones determined, for instance, by a non--trivial massive gravity
term. As a matter of principle, we can construct general inhomogeneous and locally anisotropic cosmological metrics
depending on all spacetime coordinates, see examples in Refs. \cite%
{vmgpapers}, which requests a more advanced geometric techniques. In this work, we restrict our approach only for generating off--diagonal
cosmological target metrics which can be finally transformed in "almost"
diagonalized cosmological models with dependence on a time like coordinate $%
t $ and study corresponding ekpyrotic scenarios. Such nonlinear solutions
can not be constructed if  we impose from the very beginning certain
homogeneity and symmetry conditions. This is the property of nonlinear/
parametric/ nonholonomic systems of partial derivative equations (PDEs).} The
off--diagonal, N--coefficients, are labelled $N_{i}^{a}(x^{k},y^{4}),$ where
(for this ansatz) $N_{i}^{3}=n_{i}(r,\theta )$ and $N_{i}^{4}=w_{i}(x^{k},t)+%
\mathring{w}_{i}(x^{k}).$ The data for the "primary" metric are
\begin{eqnarray}
\mathring{g}_{1}(r) &=&U^{2}-\mathring{h}_{4}(\mathring{w}_{1})^{2},%
\mathring{g}_{2}(r)=\widetilde{\alpha }^{2}r^{2},\mathring{h}_{3}=\widetilde{%
\alpha }^{2}r^{2}\sin ^{2}\theta ,\mathring{h}_{4}=\sqrt{|fU^{2}-V^{2}|},
\notag \\
\mathring{w}_{1} &=&\epsilon \sqrt{f}U^{2}/\mathring{h}_{4},\mathring{w}%
_{2}=0,\mathring{n}_{i}=0,  \label{pmdata}
\end{eqnarray}%
when the coordinate system is such way fixed that the values $f,U,V$ in (\ref%
{pgm}) result in a coefficient $\mathring{g}_{1}$ depending only on $r$.

We shall work with respect to a class of N--adapted (dual) bases
\begin{eqnarray}
\mathbf{e}_{\alpha } &=&(\mathbf{e}_{i}=\partial _{i}-N_{i}^{b}\partial
_{b},e_{a}=\partial _{a}=\partial /\partial y^{a})\mbox{ and }  \notag \\
\mathbf{e}^{\beta } &=&(e^{j}=dx^{i},\mathbf{e}^{b}=dy^{b}+N_{c}^{b}dy^{c}),
\label{nframes}
\end{eqnarray}%
which are nonholonomic (equivalently, anholonomic) because, in general,
there are satisfied relations of type $\mathbf{e}_{\alpha }\mathbf{e}_{\beta
}-\mathbf{e}_{\beta }\mathbf{e}_{\alpha }=W_{\alpha \beta }^{\gamma }\mathbf{%
e}_{\gamma },$ for certain nontrivial anholonomy coefficients $W_{\alpha
\beta }^{\gamma }(u).$ It should be noted that a nonholonomic $2+2$
splitting (\ref{nframes}) can be always defined on a pseudo--Riemannian
manifold as a fibred structure. The operators $\mathbf{e}_{i}$ and $\mathbf{e%
}^{b}$ are called respectively N--elongated partial derivatives and
N--elongated differentials because they elongate the usual ones with certain
linear N--terms. We can re--define all geometric and physical objects of a
MGT, massive, or GR, model in N--adapted form, i.e. with respect to
N--adapted bases. The motivation for such constructions is that the
gravitational and matter field equations for mentioned type theories written
for $\widehat{\mathbf{D}}$ decouple in very general form with respect to
N--adapted frames. This allows to generate exact solutions with
off--diagonal metrics and generalized connections and (if necessary) to
impose additional conditions for LC--configurations. It is not possible to
decouple such generic nonlinear systems of equations and find solutions if
we work, for instance, in coordinate frames and, from the very beginning,
with the Levi--Civita connection.\footnote{%
The N--connection geometry with nonholonomic 2+2 splitting is an example of "diadic" and/or
"tetradic/vierbein" formalism which is well known from GR
textbooks. It was applied for constructing exact solutions
with two Killing symmetries and with respect to certain special systems of reference
with rotation/ axial  symmetries. In our works
\cite{vadm1,vmgpapers}, we proved that it is possible to  decouple and solve the
gravitational field equations for a large class of gravity theories in very general forms (with coefficients of  off--diagonal metrics and generalized connections  depending on all spacetime variables). We have shown in explicit form how to define certain classes of reference and perform
N--adapted geometric constructions for an "auxiliary" canonical
d--connection $\widehat{\mathbf{D}}$, which allows to solve such sophisticate systems of nonlinear PDEs.
 At the end, it is possible to impose additional nonholonomic constraints resulting in Levi--Civita configurations. We can
not argue that this way we are able to find the "general solution" of  the Einstein equations, or any
modified/generalized versions,  because (for generic nonlinear
systems) it is not possible to prove uniqueness theorems  and other properties of solutions which are typical for linear systems. Nevertheless, we elaborated on advanced geometric methods for constructing various classes of generic off--diagonal exact and approximate solutions depending on all spacetime
coordinates via  generating functions and integration functions and constants.
Usually, the physical meaning and fundamental properties of such solutions are not known.
In certain cases of models with prescribed symmetries / topological / singularity configurations,    for "small" off--diagonal deformations with  "gravitational polarizations", inhomogeneities and local anisotropies and limits to
well--known physical/ cosmological solutions, it is possible to speculate on  nonlinear
quantum and classical effects, modified/ massive gravity  contributions, analogous models etc. }

For simplicity, we shall consider energy momentum sources (\ref{memt}) and
effective (\ref{cosmconst}) which up to frame/coordinate transforms can be
parameterized in the form {\small
\begin{equation}
\Upsilon _{\beta }^{\alpha }=\frac{1}{M_{Pl}^{2}(\partial _{\widehat{R}}%
\widehat{f})}(\mathbf{T}_{\beta }^{\alpha }+\alpha ^{-1}\mathbf{X}_{\beta
}^{\alpha })=\frac{1}{M_{Pl}^{2}(\partial _{\widehat{R}}\widehat{f})}(\
^{m}T+\alpha ^{-1})\delta _{\beta }^{\alpha }=(\ ^{m}\Upsilon +\ ^{\alpha
}\Upsilon )\delta _{\beta }^{\alpha }  \label{effectsourc}
\end{equation}%
} for constant values $\ ^{m}\Upsilon :=M_{Pl}^{-2}(\partial _{\widehat{R}}%
\widehat{f})^{-1}\ ^{m}T$ and $\ ^{\alpha }\Upsilon =M_{Pl}^{-2}(\partial _{%
\widehat{R}}\widehat{f})^{-1}\alpha ^{-1},$ with respect to N--adapted
frames (\ref{nframes}).\footnote{%
In general, such sources are not diagonal and may depend on spacetime
coordinates. We fix such N--adapted parameterizations which will allow to
construct exact solutions in explicit form.}

Let us explain an important decoupling property of the gravitational field
equations in GR and various generalizations/ modifications studied in
details in Refs. \cite{vadm1}. That anholonomic frame deformation method
(AFDM) can be applied for decoupling, and constructing solutions of the MGT
field equations (\ref{mgrfe}) with any effective source (\ref{effectsourc}).
We consider target off--diagonal metrics $\mathbf{g}=(g_{i}=\eta _{i}%
\mathring{g}_{i},h_{a}=\eta _{a}\mathring{h}_{a},N_{j}^{a})$ [there is not
summation on repeating indices in this formula] with coefficients determined
by ansatz (\ref{offdans}). For convenience, we shall use brief denotations
for partial derivatives: $\partial _{1}\psi =\psi ^{\bullet },\partial
_{2}\psi =\psi ^{\prime },\partial _{3}\psi =\psi ^{\diamond }$ and $%
\partial _{4}\psi =\psi ^{\ast }.$ Computing the N--adapted coefficients of
the Ricci and Einstein tensors (see details in Refs. \cite{vadm1}), we
transform (\ref{mgrfe}) into a system of nonlinear PDEs):%
\begin{eqnarray}
\widehat{R}_{1}^{1} &=&\widehat{R}_{2}^{2}\Longrightarrow \psi ^{\bullet
\bullet }+\psi ^{\prime \prime }=2(\ ^{m}\Upsilon +\ ^{\alpha }\Upsilon ),
\label{mgfeq1} \\
\widehat{R}_{3}^{3} &=&\widehat{R}_{4}^{4}\Longrightarrow \phi ^{\ast
}h_{3}^{\ast }=2h_{3}h_{4}(\ ^{m}\Upsilon +\ ^{\alpha }\Upsilon ),  \notag \\
\widehat{R}_{3k} &\Longrightarrow &n_{i}^{\ast \ast }+\gamma n_{i}^{\ast
}=0,\ \widehat{R}_{4k}\Longrightarrow \beta w_{i}-\alpha _{i}=0,\   \notag \\
\partial _{k}\omega &=&n_{k}\omega ^{\diamond }+w_{k}\omega ^{\ast },
\label{conf2}
\end{eqnarray}%
\begin{equation}
\mbox{for \ }\phi =\ln \left\vert \frac{h_{3}^{\ast }}{\sqrt{|h_{3}h_{4}|}}%
\right\vert ,\gamma :=(\ln \frac{|h_{3}|^{3/2}}{|h_{4}|})^{\ast },\ \ \alpha
_{i}=\frac{h_{3}^{\ast }}{2h_{3}}\partial _{i}\phi ,\ \beta =\frac{%
h_{3}^{\ast }}{2h_{3}}\phi ^{\ast },  \label{c1}
\end{equation}%
where $\Longrightarrow $ is used in order to show that certain equations
follow from respective coefficients of the Ricci tensor $\widehat{\mathbf{R}}%
_{\mu \nu }.$ In these formulas, the system of coordinates and polarization
functions are fixed for configurations with $g_{1}=g_{2}=e^{\psi (x^{k})}$
and nonzero values $\phi ^{\ast }$ and $h_{a}^{\ast }.$ The equations result
in solutions for the Levi--Civita configurations (with zero torsion) if the
coefficients of metrics are subjected to the conditions$\ $%
\begin{equation}
w_{i}^{\ast }=\mathbf{e}_{i}\ln \sqrt{|\ h_{4}|},\mathbf{e}_{i}\ln \sqrt{|\
h_{3}|}=0,\partial _{i}w_{j}=\partial _{j}w_{i}\mbox{ and }n_{i}^{\ast }=0.
\label{lccond}
\end{equation}

The system of nonlinear PDE (\ref{mgfeq1})--(\ref{lccond}) can be integrated
in general forms for any $\omega $ constrained by a system of linear first
order equations (\ref{conf2}). The explicit solutions are given by quadratic
elements%
\begin{equation}
ds^{2} =e^{\psi (x^{k})}[(dx^{1})^{2}+(dx^{2})^{2}]+\frac{\Phi ^{2}\omega
^{2}}{4\ (\ ^{m}\Upsilon +\ ^{\alpha }\Upsilon )}\mathring{h}_{3}[d\varphi
+\left( \partial _{k}\ n\right) dx^{k}]^{2}  -\frac{(\Phi ^{\ast })^{2}\omega ^{2}}{(\ ^{m}\Upsilon +\ ^{\alpha
}\Upsilon )\Phi ^{2}}\mathring{h}_{4}[dt+(\partial _{i}\ \widetilde{A}%
)dx^{i}]^{2}.  \label{nvlcmgs}
\end{equation}%
for any $\Phi =\check{\Phi},$ $(\partial _{i}\check{\Phi})^{\ast }=\partial
_{i}\check{\Phi}^{\ast }$ and $w_{i}+\mathring{w}_{i}=\partial _{i}\check{%
\Phi}/\check{\Phi}^{\ast }=\partial _{i}\ \widetilde{A}.$\footnote{%
We can construct exact solutions even such conditions are not satisfied,
i.e. the zero torsion conditions are not stated or there are given in
non--explicit form; this way, it is possible to generate off--diagonal
metrics and nonholonomically induced torsions etc, see details in \cite%
{vadm1}. There will pe presented physical arguments for what type of
generating/ integration functions and sources we have to chose in order to
construct realistic scenarios for Universe acceleration and observable dark
energy/ matter effects.} To generate new solutions we can consider arbitrary
nontrivial sources, $\ ^{m}\Upsilon +\ ^{\alpha }\Upsilon \neq 0,$ and
generating functions, $\Phi (x^{k},t):=e^{\phi }$ and $n_{k}=\partial
_{k}n(x^{i}).$ Such metrics are generic off--diagonal and can not be
diagonalized via coordinate transforms in a finite spacetime region because,
in general, the anholonomy coefficients $W_{\alpha \beta }^{\gamma }$ for (%
\ref{nframes}) are not zero (we can check by explicit computations). The
polarization $\eta $--functions for (\ref{nvlcmgs}) are computed in the form
{\small
\begin{equation}
\eta _{1}=e^{\psi }/\mathring{g}_{1},\eta _{2}=e^{\psi }/\mathring{g}%
_{2},\eta _{3}=\Phi ^{2}/4\ (\ ^{m}\Upsilon +\ ^{\alpha }\Upsilon ),\eta
_{4}=(\Phi ^{\ast })^{2}/(\ ^{m}\Upsilon +\ ^{\alpha }\Upsilon )\Phi ^{2}.
\label{polarf}
\end{equation}%
} So, prescribing any generating functions $\check{\Phi}(r,\theta ,t),$ $%
n(r,\theta ),$ $\omega (r,\theta ,\varphi ,t)$ and sources $\ ^{m}\Upsilon
,\ ^{\alpha }\Upsilon $ and then computing $\check{A}(r,\theta ,t),$ we can
transform any PG (and FLRW) metric $\mathbf{\mathring{g}}=(\mathring{g}_{i},%
\mathring{h}_{a},\mathring{w}_{i},\mathring{n}_{i})$ in MGT and/or GR into
new classes of generic off--diagonal exact solutions depending on all
spacetime coordinates. Such metrics define Einstein manifolds in GR with
effective sources $\ ^{m}\Upsilon +\ ^{\alpha }\Upsilon .$
Following such an approach, we have to study the properties of
fiducial St\"{u}ckelberg fields $\phi ^{\underline{\mu }}$ and the
corresponding bimetric structure resulting in target solutions $\mathbf{g}%
=(g_{i},h_{a},N_{j}^{a})$:

Let us analyze the "prime" configurations related
to $\mathring{\phi}^{\underline{\mu }}=(\mathring{\phi}^{\underline{i}%
}=a(\tau )\rho \widetilde{\alpha }^{-1}\widehat{n}^{\underline{i}},\mathring{%
\phi}^{\underline{3}}=a(\tau )\rho \widetilde{\alpha }^{-1}\widehat{n}^{%
\underline{3}},\mathring{\phi}^{\underline{4}}=\tau \kappa ^{-1}),$ when the
corresponding prime PG--metric $\mathbf{\mathring{g}}$ is taken in FLRW form
\begin{equation*}
ds^{2}=a^{2}(d\rho ^{2}/(1-K\rho ^{2})+\rho ^{2}(d\theta ^{2}+\sin
^{2}\theta d\varphi ^{2}))-d\tau ^{2}.
\end{equation*}
The related fiducial tensor (\ref{bm}) is computed
\begin{equation*}
\mathring{\Sigma}_{\underline{\mu }\underline{\nu }}du^{\underline{\mu }}du^{%
\underline{\nu }}=\frac{a^{2}}{\tilde{\alpha}^{2}}[d\rho ^{2}+\rho
^{2}(d\theta ^{2}+\sin ^{2}\theta d\varphi ^{2})+2H\rho d\rho d\tau -(\frac{%
\tilde{\alpha}^{2}}{\kappa ^{2}a^{2}}-H^{2}\rho ^{2})d\tau ^{2}],
\end{equation*}%
where the coefficients and coordinates are re--defined in the form $%
r\rightarrow \rho =\widetilde{\alpha }r/a(\tau )$ and $t\rightarrow \tau
=\kappa t,$ for $K=0,\pm 1;\kappa $ is an integration constant; $H:=d\ln
a/d\tau $ and the local coordinates are parameterized in the form $x^{%
\underline{1}}=\rho ,x^{\underline{2}}=\theta ,y^{\underline{3}}=\varphi ,y^{%
\underline{4}}=\tau .$

For a target metric $\mathbf{g}={\mathbf{g}_{\alpha \beta }}$ and frames $%
\mathbf{e}_{\alpha }^{\ }=\mathbf{e}_{\alpha }^{\ \underline{\alpha }%
}\partial _{\underline{\alpha }},$ we can write
\begin{equation*}
\mathbf{g}_{\alpha \beta }=\mathbf{e}_{\alpha }^{\ \underline{\alpha }}%
\mathbf{e}_{\beta }^{\ \underline{\beta }}\eta _{\underline{\alpha }%
\underline{\beta }}=\left[
\begin{array}{cc}
g_{ij}+N_{i}^{a}N_{j}^{b}g_{ab} & N_{i}^{a}h_{ab} \\
N_{i}^{a}h_{ab} & h_{ab}%
\end{array}%
\right] ,\mbox{ for }\mathbf{e}_{\alpha }^{\ \underline{\alpha }}=\left[
\begin{array}{cc}
\mathbf{e}_{i}^{\ \underline{i}} & N_{i}^{b}\mathbf{e}_{b}^{\ \underline{a}}
\\
0 & \mathbf{e}_{a}^{\ \underline{a}}%
\end{array}%
\right] .
\end{equation*}%
The values $g_{ij}=\mathbf{e}_{i}^{\ \underline{\alpha }}\mathbf{e}_{j}^{\
\underline{\beta }}\eta _{\underline{\alpha }\underline{\beta }}=e^{\psi
}\delta _{ij}=diag[\eta _{i}\mathring{g}_{i}]$, $h_{ab}=\mathbf{e}_{i}^{\
\underline{\alpha }}\mathbf{e}_{j}^{\ \underline{\beta }}\eta _{\underline{%
\alpha }\underline{\beta }}=diag[\eta _{a}\mathring{h}_{a}]$, $%
N_{i}^{3}=\partial _{k}\ n$ and $N_{i}^{4}=\partial _{k}\ \widetilde{A}$ are
related algebraically to data (\ref{polarf}) resulting in off--diagonal
solutions (\ref{nvlcmgs}). We can compute the "target" St\"{u}ckelberg
fields as $\phi ^{\mu ^{\prime }}=\mathbf{e}_{\ \underline{\mu }}^{\mu
^{\prime }\ }\phi ^{\underline{\mu }}$ with $\mathbf{e}_{\ \underline{\mu }%
}^{\mu ^{\prime }\ }$ being inverse to $\mathbf{e}_{\alpha }^{\ \underline{%
\alpha }},$ and the fiducial tensor $\Sigma _{\alpha \beta }=(\mathbf{e}%
_{\alpha }^{\ }\phi ^{\underline{\mu }})(\mathbf{e}_{\beta }^{\ }\phi ^{%
\underline{\nu }})\eta _{\underline{\mu }\underline{\nu }}=\mathbf{e}%
_{\alpha }^{\ \underline{\alpha }}\mathbf{e}_{\beta }^{\ \underline{\beta }%
}\Sigma _{\underline{\alpha }\underline{\beta }}.$ If the value $\mathring{%
\Sigma}_{\underline{\mu }\underline{\nu }}$ carries information about two
constants $\kappa $ and $\widetilde{\alpha },$ a tensor $\Sigma _{\mu \nu }$
associated to off--diagonal solutions encodes data about generating and
integration functions and via superpositions on possible Killing symmetries,
on various integration constants \cite{geroch}. In the framework of MGT, two
cosmological solutions $\mathbf{\mathring{g}}$ and $\mathbf{g}$ related by
nonholonomic deformations\footnote{%
involving not only frame transforms but also deformation of the linear
connection structure when at the end there are imposed additional
constraints for zero torsion} are characterised respectively by two
invariants $\mathring{I}^{\underline{\alpha }\underline{\beta }}=\mathring{g}%
^{\alpha \beta }\partial _{\alpha }\mathring{\phi}^{\underline{\alpha }%
}\partial _{\beta }\mathring{\phi}^{\underline{\beta }}$ and $\mathbf{I}%
^{\alpha \beta }=\mathbf{g}^{\alpha \beta }\mathbf{e}_{\alpha }\phi ^{%
\underline{\alpha }}\mathbf{e}_{\beta }\phi ^{\underline{\beta }}.$ The
tensor $\mathring{I}^{\underline{\alpha }\underline{\beta }}$ does not
contain singularities because there are not coordinate singularities on
horizon for PG metrics. The symmetry of $\Sigma _{\mu \nu }$ is not the same
as that of $\mathring{\Sigma}_{\underline{\mu }\underline{\nu }}$ and the
singular behaviour of $\mathbf{I}^{\alpha \beta }$ depends on the class of
generating/ integration functions chosen for constructing a target
solution $\mathbf{g}$.

In GR and/or Einstein--Finsler gravity theories \cite{vadm1},
off--diagonal cosmological solutions of type (\ref{nvlcmgs}) were found to
generalize various models of Bianchi, Kasner, G\"{o}del and other universes.
For instance, Bianchi type anisotropic cosmological metrics are generated if
we impose corresponding Lie algebra symmetries on metrics. It was emphasized
in \cite{kobayashi} that "any PG--type solution in general relativity (with
a cosmological constant) is also a solution to massive gravity." Such a
conclusion can be extended to a large class of generic off--diagonal
cosmological solutions generated by effective cosmological constants but it
is not true, for instance, if we consider nonholonomic deformations with
nonholonomically induced torsion like in metric compatible Finsler theories.

We note that the analysis of cosmological perturbations around an
off--diagonal cosmological background is not trivial because the fiducial
and reference metrics do not respect the same symmetries. Nevertheless,
fluctuations around de Sitter backgrounds seem to have a decoupling limit
which implies that one can avoid potential ghost instabilities if the
parameter choice is considered both for diagonal and off--diagonal
cosmological solutions, see details in \cite{fluct}. This special choice
also allows us to have a structure $X_{\mu \nu }\sim g_{\mu \nu }$ at list
in N--adapted frames when the massive gravity effects can be approximated by
effective cosmological constants and exact solutions in MGT which are also
solutions in GR.

Let us consider three examples of off--diagonal cosmological solutons with
solitonic modifications in MGT and (with alternative interpretation) GR. Two
and three dimensional solitonic waves are typical nonlinear wave
configurations which can be used for generating spacetime metrics with
Killing, or non--Killing, symmetries characterised by additional parametric
dependencies and solitonic symmetries.

\vskip5pt

\textbf{Example 1:} Taking a nonlinear radial (solitonic, with left $s$%
-label) generating function
\begin{equation}
\Phi =\ ^{s}\check{\Phi}(r,t)=4\arctan e^{q\sigma (r-vt)+q_{0}}
\label{1solgf}
\end{equation}%
and $\omega =1,$ we construct a metric a metric of type (\ref{nvlcmgs}),
\begin{equation}
\mathbf{ds}^{2} =e^{\psi (r,\theta )}(dr^{2}+\ d\theta ^{2})+~\frac{\ ^{s}%
\check{\Phi}^{2}}{4\ (\ ^{m}\Upsilon +\ ^{\alpha }\Upsilon )}\mathring{h}%
_{3}(r,\theta )d\varphi ^{2} -\frac{(\partial _{t}\ ^{s}\check{\Phi})^{2}}{%
(\ ^{m}\Upsilon +\ ^{\alpha }\Upsilon )\ ^{s}\check{\Phi}^{2}}\mathring{h}%
_{4}(r,t)[dt+(\partial _{r}\ \widetilde{A})dr]^{2},  \label{offdsol1}
\end{equation}%
where, for simplicity, we consider $n(r,\theta )=0,$\ $\widetilde{A}(r,t)$
is defined as a solution of $\ ^{s}\check{\Phi}^{\bullet }/\ ^{s}\check{\Phi}%
^{\ast }=\partial _{r}\ \widetilde{A}$ and $\mathring{h}_{a}$ are given by
PG--data (\ref{pmdata}). The generating function (\ref{1solgf}), where $%
\sigma ^{2}=(1-v^{2})^{-1}$ for constants $q,q_{0},v,$ is just a 1--soliton
solution of the sine--Gordon equation $\ ^{s}\check{\Phi}^{\ast \ast }-\ ^{s}%
\check{\Phi}^{\bullet \bullet }+\sin \ ^{s}\check{\Phi}=0$. For any class of
small polarizations with $\eta _{a}\sim 1),$ we can consider that the source
$(\ ^{m}\Upsilon +\ ^{\alpha }\Upsilon )$ is polarized by $\ ^{s}\check{\Phi}%
^{-2}$ when $h_{3}\sim \mathring{h}_{3}$ and $h_{4}\sim \mathring{h}_{4}(\
^{s}\check{\Phi}^{\ast })^{2}/\ ^{s}\check{\Phi}^{-4}$ with an off--diagonal
term $\partial _{r}\ \widetilde{A}$ resulting in a stationary solitonic
modification of the PG universe. If we chose a generating function that $%
(\partial _{\widehat{R}}\widehat{f})^{-1}=\ ^{s}\check{\Phi}^{-2}$ for the
source (\ref{effectsourc}), we can model off--diagonal \ gravitational
interactions and "gravitational polarizations" in $\widehat{f}$--gravity of
type (\ref{act}).

In absence of matter, $\ ^{m}\Upsilon =0,$ the off--diagonal cosmology is
completely determined by $\ ^{\alpha }\Upsilon $ when the nonlinear
solitonic generating function $\ ^{s}\check{\Phi}$ transform $\mathring{\mu}$
into an anisotropically polarized/variable mass of solitonic waves. Such
configurations can be modelled if $\ ^{m}\Upsilon \ll \ ^{\alpha }\Upsilon .$

If $\ ^{m}\Upsilon \gg \ ^{\alpha }\Upsilon ,$ the above soltuon describes
cosmological models determined by respective distributions of matter fields
when contributions from massive gravity are with small anisotropic
polarization. We approximate PG--metrics of type (\ref{pgm}) for a class of
nonholonomic constraints on $\Phi $ and $\psi $ (which may be not of
solitonic type) when the solutions (\ref{offdsol1}) are of type (\ref%
{offdans}) with $\eta _{\alpha }\sim 1$ and $n_{i},w_{i}\sim 0.$

Hence, by an appropriate choice of generating functions and sources, we can
model equivalently modified gravity effects, massive gravity contributions
or matter field configurations and/or MGT interactions. Such configurations
can be modeled alternatively in the framework of some classes of
off--diagonal solutions in Einstein gravity with effective cosmological
constants and respective generating functions and integration constants.

\vskip5pt

\textbf{Example 2:} Three dimensional solitonic anistoropic waves can be
generated, for instance, if we take instead of (\ref{1solgf}) a generating
functions $\ ^{s}\check{\Phi}(r,\theta ,t)$ which is a solution of the
Kadomtsev--Petivashvili, KdP, equations \cite{kadom},%
\begin{equation*}
\pm \ ^{s}\check{\Phi}^{\prime \prime }+(\ ^{s}\check{\Phi}^{\ast }+\ ^{s}%
\check{\Phi}\ \ ^{s}\check{\Phi}^{\bullet }+\epsilon \ ^{s}\check{\Phi}%
^{\bullet \bullet \bullet })^{\bullet }=0,
\end{equation*}%
when solutions induce certain anisotropy on $\theta .$\footnote{%
In a similar form, we can construct various types of vacuum gravitational
2-d and 3-d configurations characterized by solitonic hierarchies and
related bi--Hamilton structures, for instance, of KdP equations with
possible mixtures with solutions for 2-d and 3-d sine--Gordon equations etc,
see details in Ref. \cite{vacarsolitonhier}.} In the dispersionless limit $%
\epsilon \rightarrow 0,$ we can consider that the solutions are independent
on $\theta $ and determined by Burgers' equation $\ ^{s}\check{\Phi}^{\ast
}+\ ^{s}\check{\Phi}\ \ ^{s}\check{\Phi}^{\bullet }=0.$ The solutions can be
parameterized and treated similarly to (\ref{offdsol1}) but with, in
general, a nontrivial term $(\partial _{\theta }\ \widetilde{A})d\theta $
after $\mathring{h}_{4},$ when $\ ^{s}\check{\Phi}^{\bullet }/\ ^{s}\check{%
\Phi}^{\ast }=\ \widetilde{A}^{\bullet }$ and $\ ^{s}\check{\Phi}^{\prime
}/\ ^{s}\check{\Phi}^{\ast }=\ \widetilde{A}^{\prime }.$

\vskip5pt

\textbf{Example 3:}\ Choosing respective types of generating functions, we
can construct different classes of nonlinear solitonic modifications of
certain "primary" cosmological metrics with spherical symmetry which were
deformed into locally anisoropic configurations by any $\check{\Phi}%
(r,\theta ,t)$ and, finally, containing 3--d propagating solitonic waves.
For instance, such solitonic waves can be considered for a nontrivial
vertical conformal $v$--factor as in (\ref{offdans}), for instance, of KdP
type, when $\omega =\check{\omega}(r,\varphi ,t),$ when $x^{1}=r,$ $%
x^{2}=\theta ,y^{3}=\varphi ,y^{4}=t,$ for
\begin{equation}
\pm \check{\omega}^{\diamond \diamond }+(\partial _{t}\check{\omega}+\check{%
\omega}\ \check{\omega}^{\bullet }+\epsilon \check{\omega}^{\bullet \bullet
\bullet })^{\bullet }=0,  \label{kdp1}
\end{equation}%
In the dispersionless limit $\epsilon \rightarrow 0,$ the solutions are
independent on angle $\varphi $ and determined by Burgers' equation $\check{%
\omega}^{\ast }+\check{\omega}\ \check{\omega}^{\bullet }=0.$ The conditions
(\ref{conf2}) impose an additional constraint
\begin{equation*}
\mathbf{e}_{1}\check{\omega}=\check{\omega}^{\bullet }+w_{1}(r,\theta
,\varphi )\check{\omega}^{\ast }+n_{1}(r,\theta )\check{\omega}^{\diamond
}=0.
\end{equation*}%
In the system of coordinates when $\check{\omega}^{\prime }=0,$ we can fix $%
w_{2}=0$ and $n_{2}=0.$ For any arbitrary generating function with
LC--configuration, $\check{\Phi}(r,\theta ,t),$ we construct exact solutions
\begin{equation}
\mathbf{ds}^{2}=e^{\psi (r,\theta )}(dr^{2}+\ d\theta ^{2})+~\frac{\check{%
\Phi}^{2}\check{\omega}^{2}}{4\ (\ ^{m}\Upsilon +\ ^{\alpha }\Upsilon )}%
\mathring{h}_{3}(r,\theta )d\varphi ^{2}-\frac{(\partial _{t}\ \check{\Phi}%
)^{2}\check{\omega}^{2}}{(\ ^{m}\Upsilon +\ ^{\alpha }\Upsilon )\ \check{\Phi%
}^{2}}\mathring{h}_{4}(r,t)[dt+(\partial _{r}\ \widetilde{A})dr]^{2},
\label{offdsol3}
\end{equation}%
which are generic off--diagonal and depend on all spacetime coordinates.
Such cosmological solutions are with polarizations on two angles $\theta $
and $\varphi .$ Nevertheless, the character of anisotropies is different for
metrics of type (\ref{offdsol1}) and (\ref{offdsol3}). In the third class of
metrics, we obtain a Killing symmetry on $\partial _{\varphi }$ only in the
limit $\check{\omega}\rightarrow 1,$ but in the first two ones, such a
symmetry exists generically. For (\ref{offdsol3}), the value $\check{\Phi}$
is not obligatory a solitonic one which can be used for additional
off--diagonal modifications of solutions and various types of polarizations.
We can provide an interpretation similar to that in Example 1, if the
generating and integration functions are chosen to satisfy the conditions $%
\eta _{\alpha }\sim 1$ and $n_{i},w_{i}\sim 0,$ we approximate PG--metrics
of type (\ref{pgm}). In a particular case, we can use a conformal $v$%
--factor which is a 1--solitonic one, i.e. $\check{\omega}\rightarrow \omega
(r,t)=4\arctan e^{q\sigma (r-vt)+q_{0}},$ where $\sigma ^{2}=(1-v^{2})^{-1}$
and constants $q,q_{0},v,$ defines a 1--soliton solution of the sine--Gordon
equation $\omega ^{\ast \ast }-\omega ^{\bullet \bullet }+\sin \omega =0.$
Such a soliton propagates in time along the radial coordinate.

The solitonic waves constructed for Examples 1-3 can be characterized by
corresponding velocities and effective mass and energy. They may model
different dark energy and polarized dark matter distributions and nonliner
effects determined by certain source tems $\ ^{m}\Upsilon :=\
^{m}T/M_{Pl}^{2}\partial _{\widehat{R}}\widehat{f}$ and $\ ^{\alpha
}\Upsilon =1/M_{Pl}^{2}\alpha (\partial _{\widehat{R}}\widehat{f})$ in MGT
defined by (\ref{effectsourc}).

We now consider a reconstruction mechanism with distinguished off--diagonal
cosmological effects \cite{vadm1} by generalizing some methods elaborated
for $f(R)$ gravity in \cite{odints}. Any cosmological solution in massive,
MGT and/or GR parameterized in a form (\ref{offdans}) (in particular, as (%
\ref{offdsol1}) and (\ref{offdsol3})) can be encoded into an effective
functional $\widehat{f}-\frac{\mathring{\mu}^{2}}{4}\mathcal{U}=f(\widehat{R}%
),\widehat{R}_{\mid \widehat{\mathbf{D}}\rightarrow \nabla }=R$ (\ref%
{mgrfunct}). This allows us to work as in MGT; the conditions $\partial
_{\alpha }f(\widehat{R})=0$ if $\widehat{R}=const$ simplify substantially
the computations. The starting point is to consider a prime flat FLRW like
metric
\begin{equation*}
ds^{2}=a^{2}(t)[(dx^{1})^{2}+(dx^{2})^{2}+(dy^{3})^{2}]-dt^{2},
\end{equation*}%
where $t$ is the cosmological time. In order to extract a monotonically
expanding and periodic cosmological scenario, we parameterize $\ln
|a(t)|=H_{0}t+\tilde{a}(t)$ for a periodic function $\tilde{a}(t+\tau )=\
^{1}a\cos (2\pi t/\tau ),$ where $0<\ ^{1}a<H_{0}.$ Our goal is to prove
that such a behavior is encoded into off--diagonal solutions of type (\ref%
{offdsol1})--(\ref{offdsol3}).

We write FLRW like equations with respect to N--adapted (moving) frames (\ref%
{nframes}) for a generalized Hubble function $H,$%
\begin{equation*}
3H^{2}=8\pi \rho \mbox{ and }3H^{2}+2\mathbf{e}_{4}H=-8\pi p.
\end{equation*}%
Using variables with $\partial _{\alpha }f(\widehat{R})_{|\widehat{R}%
=const}=0$, we can consider a function $H(t)$ when $\mathbf{e}_{4}H=\partial
_{t}H=H^{\ast }.$ The energy--density and pressure of an effective perfect
fluid are computed
\begin{eqnarray}
\rho &=&(8\pi )^{-1}[(\partial _{R}f)^{-1}(\frac{1}{2}f(R)+3H\mathbf{e}%
_{4}(\partial _{R}f))-3\mathbf{e}_{4}H]  \label{rhomg} \\
&=&(8\pi )^{-1}[\partial _{\widehat{R}}\ln \sqrt{|\widehat{f}|}-3H^{\ast
}]=(8\pi )^{-1}[\partial _{\widehat{R}}\ln \sqrt{|\frac{\mathring{\mu}^{2}}{4%
}\mathcal{U}+f(\widehat{R})|}-3H^{\ast }],  \notag \\
p &=&-(8\pi )^{-1}[(\partial _{R}f)^{-1}(\frac{1}{2}f(R)+2H\mathbf{e}%
_{4}(\partial _{R}f)+\mathbf{e}_{4}\mathbf{e}_{4}(\partial _{R}f))+\mathbf{e}%
_{4}H]  \notag \\
&=&(8\pi )^{-1}[\partial _{\widehat{R}}\ln \sqrt{|\widehat{f}|}+H^{\ast
}]=-(8\pi )^{-1}[\partial _{\widehat{R}}\ln \sqrt{|\frac{\mathring{\mu}^{2}}{%
4}\mathcal{U}+f(\widehat{R})|}+H^{\ast }].  \notag
\end{eqnarray}%
In N--adapted variables, the equation of state, EoS, parameter for the
effective dark fluid is defined by
\begin{equation}
w=\frac{p}{\rho }=\frac{\widehat{f}+2H^{\ast }\partial _{\widehat{R}}%
\widehat{f}}{\widehat{f}-6H^{\ast }\partial _{\widehat{R}}\widehat{f}}=\frac{%
\frac{\mathring{\mu}^{2}}{4}\mathcal{U}+f(\widehat{R})+2H^{\ast }\partial _{%
\widehat{R}}\widehat{f}}{\frac{\mathring{\mu}^{2}}{4}\mathcal{U}+f(\widehat{R%
})-6H^{\ast }\partial _{\widehat{R}}\widehat{f}},  \label{wcosm}
\end{equation}%
when the corresponding EoS is $p=-\rho -(2\pi )^{-1}H^{\ast }$ and $\mathcal{%
U}(t)$ is computed, for simplicity, for a configuration of "target" St\"{u}%
ckelberg fields $\phi ^{\mu ^{\prime }}=\mathbf{e}_{\ \underline{\mu }}^{\mu
^{\prime }\ }\phi ^{\underline{\mu }}$ when a found solution is finally
modelled by generating functions with dependencies on $t.$

Taking a generating Hubble parameter $H(t)=H_{0}t+H_{1}\sin \omega t,$ for $%
\omega =2\pi /\tau ,$ we can recover the modified action for oscillations of
off--diagonal (massive) universe (see similar details in \cite{odints}),%
\begin{equation}
f(R(t))=6\omega H_{1}\int dt[\omega \sin \omega t-4\cos \omega
t(H_{0}+H_{1}\sin \omega t)]\exp [H_{0}t+\frac{H_{1}}{\omega}\sin \omega t].
\label{recovf}
\end{equation}%
We can not invert analytically to find in explicit form $R.$ Nevertheless,
we can prescribe any values of constants $H_{0}$ and $H_{1}$ and of $\omega $
and compute effective dark energy and dark matter oscillating cosmology
effects for any off--diagonal solution in massive gravity and/or effective
MGT, GR. To extract contributions of $\mathring{\mu}$ we can fix, for
instance, $\widehat{f}(\widehat{R})=$ $\widehat{R}=R$ and using (\ref{act})
and (\ref{act1}) we can relate $f(R(t))$ and respective constants to certain
observable data in cosmology.

The MGT theories studied in this work encode, for respective nonholonomic
constraints, the ekpyrotic scenario which can be modelled similarly to $f(R)
$ gravity. A scalar field is introduced into usual ekpyrotic models in order
to reproduce a cyclic universe and such a property exists if we consider
off--diagonal solutions with massive gravity terms and/or $f$%
--modifications. Let us consider a prime configuration with energy--density
for pressureless matter $\mathring{\rho}_{m},$ for radiation and
anisotropies we take respectively $\mathring{\rho}_{r}$ and $\mathring{\rho}%
_{\sigma }$ for radiation and anisotropies, $\kappa $ is the spatial
curvature of the universe and a target effective energy--density $\rho $ (%
\ref{rhomg}). A FLRW model can be described by
\begin{equation*}
3H^{2}=8\pi \lbrack \frac{\mathring{\rho}_{m}}{a^{3}}+\frac{\mathring{\rho}%
_{r}}{a^{4}}+\frac{\mathring{\rho}_{\sigma }}{a^{6}}-\frac{\kappa }{a^{2}}%
+\rho ].
\end{equation*}%
We generate an off--diagonal/massive gravity cosmological \ cyclic scenario
containing a contracting phase by solving the initial problems if $w>1,$ see
(\ref{wcosm}). A homogeneous and isotripic spatially flat universe is
obtained when the scale factor tends to zero and the effective $f$--terms
(massive gravity and off--diagonal contributions) dominate over the rests.
In such cases, the results are similar to those in the inflationary
scenario. For recovering (\ref{recovf}), the ekpyrotic scenario takes place
and mimic the observable universe for $t\sim \pi /2\omega $ in the effective
EoS parameter $w\approx -1+\sin \omega t/3\omega H_{1}\cos ^{2}\omega t\gg 1
$. This allows us to conclude that in massive gravity and/or using
off--diagonal interactions in GR cyclic universes can be reconstructed in
such forms that the initial, flatness and/or horizon problems can be solved.

In the diversity of off--diagonal cosmological solutions which can
constructed using above presented methods, there are cyclic ones with
singularities of the type of big bang/ crunch behaviour. Choosing necessary
types generating and integration functions, we can avoid singularities and
elaborate models with smooth transition. Using the possibility to generate
nonholonomically constrained $f$--models with equivalence to certain classes
of solutions in massive gravity and/or off--diagonal configurations in GR,
we can study in this context, following methods in \cite{odints,vadm1}, big
and/or little rip cosmology models, when the phantom energy--density is
modelled by off-diagonal interactions. We omit such considerations in this
letter.

In summary, we have found new cosmological off--diagonal solutions in
massive gravity with flat, open and closed spatial geometries. We applied a
geometric techniques for decoupling the field equations and constructing
exact solutions in $f(R)$ gravity, theories with nontrivial torsion and
noholonomic constraints to GR and possible extensions on (co) tangent
Lorentz bundles. A very important property of such generalized classes of
solutions is that they depend, in principle, on all spacetime coordinates
via generating and integration functions and constants. After some classes
of solutions were constructed in general form, we can impose at the end
nonholonomic constraints, cosmological approximations, extract
configurations with a prescribed spacetime symmetry, consider asymptotic
conditions etc. Thus, our solutions can be used not only for elaborating
homogeneous and isotropic cosmological models with arbitrary spatial
curvature, but also for study "non--spherical" collapse models of the
formation of cosmic structure such as stars and galaxies (see also \cite%
{kobayashi}).

Elaborating cosmological scenarios only for diagonalizable metrics with
spherical and/or two Killing symmetries, there are limited possibilities to
study  nonlinear  physical effects in massive gravity theory and to determine if there are differences from similar ones in  $f$--gravity and/or GR. The solutions of corresponding nonlinear systems of ordinary differential equations are
parameterized  by integration constants with certain explicit values fixed  in order to
satisfy certain boundary/asymptotic  conditions, experimental data etc. We positively have to modify the GR
theory in order to explain observational data in modern  cosmology with acceleration and dark energy and dark matter  and to develop self--consistent  quantum models of (non) massive gravity.

In a series of our works \cite{vadm1,vmgpapers} on geometric methods and
exact solutions, we proposed  a new approach  constructing exact solutions in commutative and noncommutative MGTs, massive gravity and GR. The fundamental field equations in such  gravity models
consist very sophisticate off--diagonal coupled systems of nonlinear PDEs which in the past could  be solved in very special cases. Nevertheless, we proved that it is possible to decouple and solve  such equations in certain
general forms by considering necessary type nonholonomic transforms and
 "auxiliary" connections which can be constrained at the end in order to extract LC--configurations.  Usually, it is not clear what physical meaning may have such general off--diagonal inhomogeneous and
locally anisotropic solutions but it is obvious that they should  be important for
certain cosmological scales. A new and important feature of such solutions is that the
off--diagonal anisotropic configurations allow us to model scenarios of cosmic
acceleration and massive gravity and/or dark energy and dark matter effects
as certain nonlinear/ parametric / nonholonomic interactions in effective
Einstein spaces. Following such an "orthodox paradigm",  it is natural to put the question:\ \textit{May be it is not necessary to modify  the Einstein gravity at classical level  but  to consider new classes of off--diagonal solutions for nonlinear and nonholonomic gravitational and matter fields interactions and try to apply this in modern cosmology?} It is not possible to give a complete answer to such fundamental problems   in this letter or in other articles with explicit constructions on modeling MGTs as  generalized Einstein spaces with nontrivial vacuum structure and non--minimal coupling with matter fields.  The goal of this paper was to  construct in explicit form and study  some typical examples of generic off--diagonal solutions determined by generating and integration functions when massive gravity effects are reproduced via nonlinear solitonic wave interactions in an effective Einstein theory. We  concluded that  it is possible to model cosmological scenarios as in $f$--modified gravity theories by considering another types of  off--diagonal configurations on  nonholonomic Einstein spaces.

We also developed a reconstruction method for the massive gravity theory
which admits and effective off--diagonal interpretation in GR and $f$%
--modified gravity with cyclic and ekpyrotic universe solution. The
expansion is around the GR action if we admit a nontrivial effective
torsion. For zero torsion constraints, it is also possible to perform
off--diagonal cosmological models keeping the constructions in the framework
of the GR theory. Our results indicate that theories with massive gravitons
and off--diagonal interactions may lead to more complicated cyclic universes.
Following such an approach, the ekpyrotic (little rip) scenario can be
realized with no need to introduce an additional field (or modifying
gravity) but only in terms of massive gravity or GR. Further constructions
can be related to reconstruction scenarios of $f(R)$ and massive gravity
theories leading to little rip universe. The dark energy for little rip
models presents an example of non--singular phantom cosmology. Finally, we
note that other types of non--singular super--accelerating universe may be
also reconstructed in $f(R)$ gravity.

\vskip5pt

\textbf{Acknowledgments:\ } The work is partially supported by the Program
IDEI, PN-II-ID-PCE-2011-3-0256. The author is grateful to organizers and
participants of respective sections MG13, where some results of this paper
were presented. He thanks S. Capozziello, S. Hervik, S. Hassan, E.
Guendelman, P. Stavrinos, D. Singleton and S. Rajpoot for important
discussions, critical remarks, or substantial support for seminars. The paper was modified during the DAAD advanced research fellowship and the author is grateful to D. L\"{u}st and O. Lechtenfeld for hosting and support. 
Finally, the author thanks the referee for hard work and very important remarks which help to improve and understand better certain results provided in this paper.


\end{document}